\documentclass{emulateapj}
\usepackage{apjfonts}

\newcommand{\te}{t_{\rm E}}
\newcommand{\re}{r_{\rm E}}
\newcommand{\thetae}{\theta_{\rm E}}

\shortauthors{Chung et al.}
\shorttitle{Detecting planets in Andromeda}

\begin{document}

\title{The possibility of detecting planets in the  Andromeda Galaxy}

\author{
S.-J. Chung\altaffilmark{1}
D. Kim\altaffilmark{1},\\
The Angstrom Collaboration:
M.J. Darnley\altaffilmark{2},
J.P. Duke\altaffilmark{2},
A. Gould\altaffilmark{4},
C. Han\altaffilmark{1,5},
Y.-B. Jeon\altaffilmark{3},
E. Kerins\altaffilmark{2},
A. Newsam\altaffilmark{2} and
B.-G. Park\altaffilmark{3}
}

\altaffiltext{1}{Department of Physics, Institute for Basic Science
Research, Chungbuk National University, Chongju 361-763, Korea}
\altaffiltext{2}{Astrophysics Research Institute, Liverpool John Moores
University, Twelve Quays House, Birkenhaed, Merseyside CH41 1LD, UK}
\altaffiltext{3}{Bohyunsan Optical Astronomy Observatory, Korea Astronomy
and Space Science Institute, Youngchon 770-820, Korea}
\altaffiltext{4}{Department of Astronomy, the Ohio State University,
140 West 18th Avenue, Columbus, OH 43210}
\altaffiltext{5}{corresponding author}

%\submitted{Submitted to The Astrophysical Journal}

\begin{abstract}
The Angstrom Project is using a global network of 2m-class telescopes
to conduct a high cadence pixel microlensing survey of the bulge of the 
Andromeda Galaxy (M31), with the primary aim of constraining its underlying
bulge mass distribution and stellar mass function.  Here we investigate
the feasibility of using such a survey to detect planets in M31.  We 
estimate the efficiency of detecting signals for events induced by 
planetary systems as a function of planet/star mass ratio and separation, 
source type and background M31 surface brightness.  We find that for 
planets of a Jupiter-mass or above that are within the lensing zone 
($\sim 1 -3$~AU) detection is possible above 3~$\sigma$, with detection 
efficiencies $\sim 3\%$ for events associated with giant stars, which are 
the typical source stars of pixel-lensing surveys.  A dramatic improvement 
in the efficiency of $\sim 40$ -- 60\% is expected if follow-up observations 
on an 8m telescope are made possible by a real-time alert system.
\end{abstract}

\keywords{gravitational lensing -- planets and satellite: general --
galaxies: individual (M31)}

\section{Introduction}

Various techniques are being used to search for extrasolar planets,
including the radial velocity technique \citep{mayor95, marcy96}, transit
method \citep{struve52}, direct imaging \citep{angel94, stahl95}, pulsar 
timing \citep{wolszczan92}, and microlensing \citep{mao91, gould92}.  See 
the reviews of \citet{perryman00, perryman05}.  The microlensing signal 
of a planetary companion to microlens stars is a short-duration perturbation 
to the smooth standard light curve of the primary-induced lensing event 
occurring on a background source star.  Once the signal is detected and 
analyzed, it is possible to determine the planet/star mass ratio, $q$, 
and the projected planet-star separation, $s$ (normalized by the angular 
Einstein ring radius $\thetae$).  Recently, two robust microlensing 
detections of  exoplanets were reported by \citet{bond04} and 
\citet{udalski05}.

The microlensing technique has various advantages over other methods.
First, microlensing is sensitive to lower-mass planets than most other
methods (except pulsar timing) and it is possible, in principle, 
to detect Earth-mass planets from ground-based observations \citep{gould04}.  
Second, the microlensing technique is most sensitive to planets located in 
the so-called lensing zone corresponding to the range of 0.6 -- 1.6 Einstein 
ring radii.  The typical value of the Einstein radius, $\re$, for Galactic 
lensing events is a couple of AU, and thus the lensing zone roughly overlaps 
with the habitable zone.  Third, the microlensing technique is the only 
proposed method that can detect and characterize free-floating planets 
\citep{bennett02, han05}.  Fourth, the biases in the search technique are 
less severe and can be quantified easily compared to other methods 
\citep{gaudi02}. Therefore, the microlensing technique will be able to 
provide the best statistics of the Galactic population of planets.

In addition to the advantages mentioned above, the microlensing technique
is distinguished from other techniques in the sense that the planets
to which it is sensitive are much more distant than those found with other 
techniques.  With the advent of photometry techniques like the difference 
imaging \citep{alard98} and pixel method \citep{melchior99}, microlensing 
searches are not restricted to the field within the Galaxy and  can be 
extended to unresolved star fields of nearby galaxies such as M31.  
Therefore, microlensing is the only feasible technique that can detect 
planets located in other galaxies.

Microlensing searches toward M31 have been and are being carried out by 
various collaborations including the POINT-AGAPE \citep{auriere01,
paulinhendriksson02, paulinhendriksson03, an04, belokurov05}, AGAPE
\citep{ansari97, ansari99}, VATT-Colombia \citep{crotts96, uglesich04},
MEGA \citep{dejong04}, and WeCAPP \citep{riffeser01, riffeser03} 
collaborations, as well as MDM \citep{calchinovati03}, McGraw-Hill 
\citep{calchinovati02}, and Nainital \citep{joshi03, joshi05} surveys.  
The monitoring frequencies of these experiments are typically $\sim 3$ 
observations per week, too low to detect planetary signals.  However, 
with the expansion of the global telescope network, the monitoring 
frequency of such surveys is rapidly increasing.  For example, a new M31 
pixel-lensing survey, the Andromeda Galaxy Stellar Robotic Microlensing 
(Angstrom) project is expected to achieve a monitoring frequency of $\sim 5$ 
observations per 24-hour period by using a network of telescopes, including 
the robotic 2m Liverpool Telescope at La Palma, Faulkes Telescope North in 
Hawaii, 1.8m telescope at the Bohyunsan Observatory in Korea, and the 2.4m 
Hiltner Telescope at the MDM Observatory in Arizona \citep{kerins05}.

The possibility of detecting planetary microlensing events caused by a
lens located in M31 was discussed by \citet{baltz01}.  However,
the main focus of that paper was evaluating the detectability of events
caused by binary lenses in general and the comment about the planetary
lensing was brief treating the planetary system as one case of the binary 
lenses.  In addition, their detection rate estimate of the M31 planetary 
lensing events was based only on events that exhibit caustic crossings, 
while a significant fraction of events with detectable planetary signals 
might be non-caustic-crossing events.  Moreover, the aim of their work 
is was the rough evaluation of feasibility and thus not based on specific 
observational setup and instruments. Similarly, the work of \citet{covone00} 
was also based on an arbitrary observational setup.

In this paper, we explore the feasibility of detecting planets in M31
from a high-frequency pixel-lensing survey using a global network of
2m-class telescopes.  The paper is organized as follows.  In \S\ 2, we
briefly describe the basics of planetary lensing.  In \S\ 3, we estimate
the efficiency of detecting planetary signals for events induced by
planetary systems with various planet-star separations and mass ratios, 
associated with source stars of different types, and occurring toward 
fields with a range of surface brightness $\mu$.  From the dependence of 
the detection efficiency on these parameters, we investigate possible 
types of detectable planets, the optimal source stars, fields, and 
observation strategy for M31 planet detections.  In \S\ 4, we discuss 
about methods to further improve the planet detection efficiency.  We 
summarize the results and conclude in \S\ 5.

\section{Basics of Planetary Lensing}

Planetary lensing is described by the formalism of a binary lens with
a very low-mass companion.  Because of the very small mass ratio, the
planetary lensing behavior is well described by that of a single lens
of the primary star for most of the event duration.  However, a
short-duration perturbation can occur when the source star passes the
region near a caustic, which represent the set of source positions
at which the magnification of a point source becomes infinite.  The
caustics of binary lensing form a single or multiple closed figures
where each figure is composed of concave curves (fold caustics) that
meet at cusps.

For the planetary case, there exist two sets of disconnected caustics.
One `central caustic' is located close to the host star.  The other
`planetary caustic' is located away from the host star and its number
is one or two depending on whether the planet lies outside ($s>1$) or 
inside ($s<1$) the Einstein ring.  The size of the caustic, which is 
directly proportional to the planet detection efficiency, is maximized 
when the planet is located in the `lensing zone', which represents the 
range of the star-planet separation of $0.6\lesssim s\lesssim 1.6$ 
\citep{gould92}.

The planetary perturbation induced by the central caustic is of special
interest for M31 pixel-lensing events.  While the perturbation induced
by the planetary caustic can occur at any part of the light curve of 
any event, even those of low magnification, the perturbation induced by 
the central caustic always occurs near the peak of the light curve of a
high-magnification event.  Then, the chance for the M31 pixel-lensing
events to be perturbed by the central caustic can be high because these 
events tend to have high magnifications.  In addition, the chance of 
detecting planetary signals for these events becomes even higher due to 
the improved photometric precision thanks to the enhanced brightness of 
the lensed source star during the time of perturbation.

\section{Detection Efficiency}

To estimate the efficiency of detecting planetary signals of M31 events,
we compute the `detectability' defined as the ratio of the planetary
signal, $\epsilon$, to the photometric precision, $\sigma_{\rm ph}$,
i.e.,
\begin{equation}
{\cal D}= {|\epsilon|\over \sigma_{\rm ph}}.
\label{eq3.1}
\end{equation}
The planetary signal is the deviation of the lensing light
curve from that of the single lensing event of the primary lens star,
and thus it is defined as
\begin{equation}
\epsilon = {{A-A_0}\over A_0},
\label{eq3.2}
\end{equation}
where $A$ is the magnification of the planetary lensing and $A_0$ is the
single lensing magnification caused by the host star alone.  For an M31
pixel-lensing event, the lensing signal is the flux variation measured
on the subtracted image, while the noise is dominated by the background
flux.  Then, the photometric precision can be approximated as
\begin{equation}
\sigma_{\rm ph} = {\sqrt{F_{\rm B}}\over F_S(A-1)},
\label{eq3.3}
\end{equation}
where $F_S$ and $F_{\rm B}$ are the baseline flux of the lensed source
star and the blended background flux, respectively.  Under this definition
of the detectability, ${\cal D}=1$ implies that the planetary signal is
equivalent to the photometric precision.

% Figure 1 ------------------------------------------------------
\begin{figure*}[t]
\epsscale{1.0}
%\plotone{f1.eps}
\caption{\label{fig:one}
Contour maps of the detectability of the planetary lensing signal, ${\cal D}$,
as a function of the source star position for events caused by planetary 
systems with various star-planet separations, $s$, and planet/star mass 
ratios, $q$, and associated with a giant source star.  The source star is 
assumed to have an absolute magnitude $M_I=0.0$ and a radius $R_\star=10.0
\ R_\odot$.  The maps are centered at the position of the primary lens 
star, and the planet is located on the left.  The contours are drawn at the 
levels of ${\cal D}=1.0$ (white), 2.0 (yellow), and 3.0 (brown), respectively. 
The white dotted circle has a radius of $u_{0,{\rm th}}$, which represents 
the threshold impact parameter of the source trajectory required for the 
event detection.  All lengths are in units of $u_{0,{\rm th}}$, whose 
absolute value is marked in the bottom left panel.  The closed figure(s) 
drawn by a blue curve is (are) the caustic(s).  The field surface brightness 
is assumed to be $\mu=18.0\ {\rm mag}/ {\rm arcsec}^2$, which is a 
representative value of the M31 bulge region.  A sample source trajectory
is indicated by the white arrow and the corresponding lightcurve is shown in 
Fig.~\ref{fig:five}.
}\end{figure*}

% Figure 2 ------------------------------------------------------
\begin{figure*}[htb]
\epsscale{1.0}
%\plotone{f2.eps}
\caption{\label{fig:two}
Same as Fig.~\ref{fig:one} except that the source star is a A5 main-sequence
with an absolute magnitude $M_I=1.73$ and a radius $R_\star=1.7\ R_\odot$.  
}\end{figure*}

% Figure 3 ------------------------------------------------------
\begin{figure*}[htb]
\epsscale{1.0}
%\plotone{f3.eps}
\caption{\label{fig:three}
Same as Fig.~\ref{fig:one} except that the source star is a F5 main-sequence
with an absolute magnitude $M_I=2.86$ and a radius $R_\star= 1.3\
R_\odot$.
}\end{figure*}

We estimate the detection efficiency for a representative event that is
most probable under the assumption that the M31 halo is not significantly
populated with MACHOs.  Under this assumption, it is expected that the
events to be detected toward the field in and around the M31 bulge,
where the event rate is highest, are caused mostly by low-mass stars
located in the bulge itself \citep{kerins05}.  We, therefore, choose
a representative event as the one caused by a lens with a primary star
mass of $m=0.3\ M_\odot$ and distances to the source star and lens of
$D_S=780$ kpc and $D_L=(780-1)$ kpc, respectively.  Then, the corresponding
physical and angular Einstein ring radii are $\re=1.56\ {\rm AU}$ and
$\thetae=2.0\ \mu{\rm as}$, respectively.  The assumed
timescale is $\te=20$ days.

To see the dependence of the detection efficiency on the stellar type
of the source star, we test several types of source stars with various
absolute magnitudes and sizes.  The type of the source star affects the
detection efficiency in two different ways.  On one side, high-luminosity
of the bright source star contributes to the detection efficiency in a
positive way because the photometric precision improves with the increase
of the source star brightness.  If the source star is too bright, on the
other hand, it is likely to be a giant star, for which the planetary
signal $\epsilon$ might be diminished due to the finite-source effect
\citep{bennett96}.  The stellar types of the tested source stars are giant,
A5, and F5 main-sequence (MS) stars with the absolute magnitudes of $M_I=0.0$,
1.73, and 2.86, and stellar radii of $R_\star=10.0\ R_\odot$, $1.7\ R_\odot$,
and $1.3\ R_\odot$, respectively.  The corresponding source angular radii
normalized by the Einstein radius are $\rho_\star=\theta_\star/\thetae=
(R_\star/D_S)/ \thetae=0.03$, 0.005, and 0.004, respectively.

Observations and photometry are assumed to be carried out as follows.
Following the specification of the Liverpool Telescope, we assume that
the instrument can detect 1 photon/s for an $I=24.2$ star.  We also
assume that the average seeing is $\theta_{\rm see}=1''\hskip-2pt .0$
and the observation is carried out such that small-exposure images are
combined to make a 30 min exposure image to obtain a high signal-to-noise
ratio while preventing saturation in the central bulge region.  The 
photometry is done such that the flux variation is measured at an aperture 
that maximizes the signal-to-noise ratio of the measured flux variation.
In the background-dominated regime such as the M31 field, the noise is
proportional to the aperture radius $\theta_{\rm ap}$, i.e.\ $F_B\propto
\pi \theta_{\rm ap}^2$.  On the other hand, assuming a gaussian PSF,
the measured source flux variation scales as $F=F_S(A-1)\propto
\int_0^{\theta_{\rm ap}}(\theta/\sigma_{\rm PSF}^2)\exp(-\theta^2/2
\sigma_{\rm PSF}^2)d\theta$, where $\sigma_{\rm PSF}=0.425\theta_{\rm see}$.
Therefore, the signal-to-noise ratio scales as $S/N =F/\sqrt{F_B}\propto
(1/\theta_{\rm ap})\int_0^{\theta_{\rm ap}}(\theta/\sigma_{\rm PSF}^2)
\exp\left( -{\theta^2/2\sigma_{\rm PSF}^2} \right) d\theta$.  Then, the
optimal aperture that maximizes the signal-to-noise ratio is
$\theta_{\rm ap}=0.673 \theta_{\rm see}$.  With the adoption of this
aperture, the fraction of the source flux within the aperture is
$F(\theta\leq \theta_{\rm ap})/F_{\rm tot}= 0.715$, where $F_{\rm tot}$
is the flux measured at $\theta_{\rm ap}\equiv \infty$.

\begin{figure}[t]
\epsscale{1.2}
\plotone{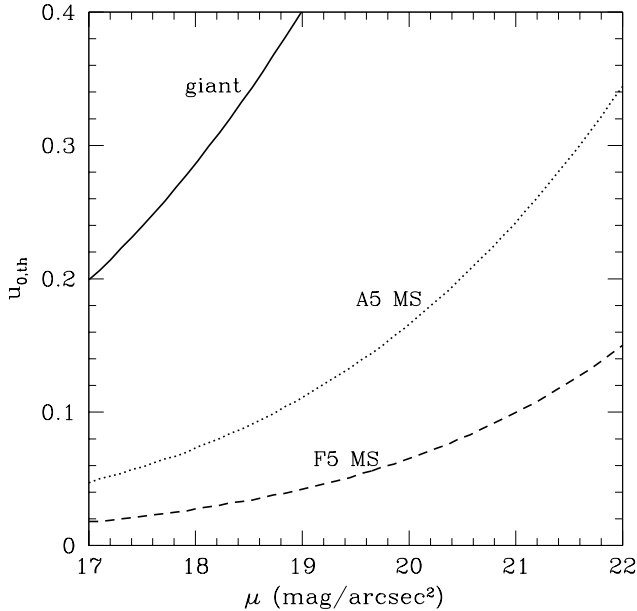}
\caption{\label{fig:four}
Threshold lens-source impact parameter, $u_{0,{\rm th}}$, as a function of 
the background surface brightness $\mu$ for source stars of various types.
}\end{figure}

In Figure~\ref{fig:one}--\ref{fig:three}, we present the contour maps of
the detectability of the planetary lensing signal as a function of the 
source star position for events caused by planetary systems with various 
$s$ and $q$, and involved with source stars of various stellar types.  
The contours are drawn at the levels of ${\cal D}=1.0$ (white), 2.0 (yellow), 
and 3.0 (brown), respectively.  We assume that the planetary signal is 
firmly detected if ${\cal D}\geq 3.0$.  In the map, we present only the 
region around the `detection zone', which represents the region of the 
source star position where the magnification is higher than a threshold 
magnification required for the event detection, $A_{\rm th}$.  The threshold 
magnification is defined by $(A_{\rm th}-1)F_S=3\sqrt{F_B}$, i.e., $3\sigma$ 
detection of the event.  If we define $u_{0,{\rm th}}$ as the threshold 
lens-source impact parameter corresponding to $A_{\rm th}$, everything of 
interest is contained within the circle with the radius $u_{0,{\rm th}}$ 
(marked by a white dotted circle in each panel).  We, therefore, use 
$u_{0,{\rm th}}$ as a scale length instead of the Einstein radius.  
However, to provide the relative size of the detection zone, we mark the 
absolute value of $u_{0,{\rm th}}$ in the bottom left panel of each figure.  
The value of the threshold impact parameter decreases as either the source 
star becomes fainter or the background surface brightness increases.  In 
Figure~\ref{fig:four}, we present the variation of the threshold impact 
parameter as a function of the background surface brightness for source 
stars of various types.  The maps are constructed for a common surface 
brightness of $\mu=18.0\ {\rm mag}/ {\rm arcmin}^2$, which is a representative 
value of the M31 bulge region.  For the construction of the maps, we consider 
the attenuation of the magnification caused by the finite-source effect.

From the maps, we find that although both the sizes of the event detection 
zone ($u_0<u_{0,th}$) and the region of the planetary perturbation ($D>3$) 
decrease as the source star becomes fainter, the rate of the decrease of 
the planetary perturbation region is smaller than the rate of decrease of
the event 
detection zone.  This is because the planetary perturbation is confined 
to the region around the caustic whose size does not depend on the source 
brightness.  As a result, the perturbation region occupies a greater 
fraction of the event detection zone with the decrease of the source star 
brightness.  However, this does not necessarily imply that the planet 
detection efficiency of events associated with MS stars is higher than 
that of events associated brighter giant stars.  This is because despite 
the slow rate of decrease, the perturbation region does decrease, and thus 
picking up the resulting short-duration planetary signals for events 
involved with faint MS stars requires higher monitoring frequency.

\begin{figure}[t]
\epsscale{1.2}
\plotone{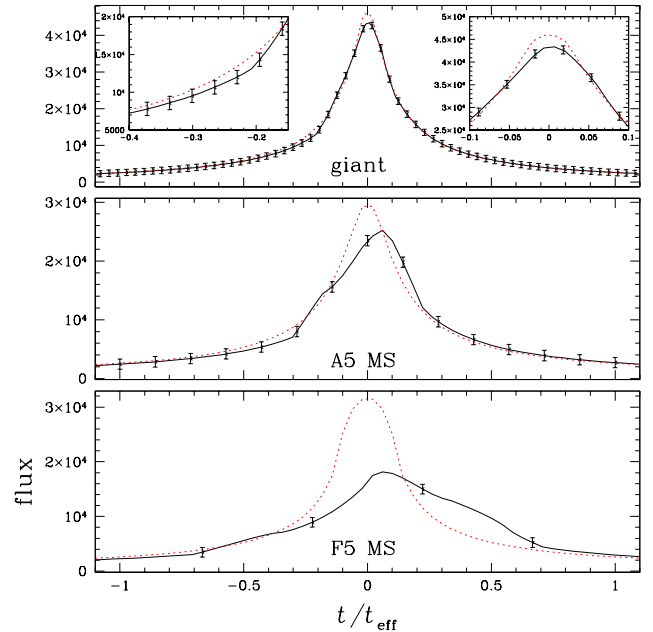}
\caption{\label{fig:five}
Example pixel-lensing light curves with planetary perturbations.  The time 
scale is normalized by the effective time scale $t_{\rm eff}$ that is the 
time required for the source to cross the threshold impact parameter.  The 
individual light curves are for events produced by a common planetary lens 
system with $s=1.2$ and $q=5\times 10^{-3}$, but occurred on different 
source types of giant, A5, and F5 main-sequence stars.  The source 
trajectories of the individual light curves are marked in the corresponding 
panels in Fig.~\ref{fig:one}, \ref{fig:two}, and \ref{fig:three}, respectively. 
The data points on each light curve are marked under the assumption that the 
monitoring frequency of 5 observations per 24-hour period.  The dotted 
curve shows the lightcurve expected in the absence of a planet.
}\end{figure}

Once the maps of the detectability are constructed, we produce a large
number of light curves of lensing events resulting from source trajectories
with random orientations and impact parameters $u_0\leq u_{0,{\rm th}}$
(see example light curves in Figure~\ref{fig:five}).
Then, we estimate the detection efficiency as the ratio of the number of 
events with detectable planetary signals to the total number of the tested 
events.  We assume that on average five combined images with a 30 min 
exposure are obtained daily following the current Angstrom survey.  By 
applying a conservative criterion for the detection of the planetary signal, 
we assume that the planet is detected if the signal with ${\cal D}\geq 3$ 
is detected at least five times during the event.  Since the monitoring 
frequency is $f=5\ {\rm times}/ {\rm day}$, this implies that the planetary 
signal should last at least 1 day for detection.

% Figure 4 ------------------------------------------------------

% Figure 5 ------------------------------------------------------

% Figure 6 ------------------------------------------------------
\begin{figure}[t]
\epsscale{1.2}
\plotone{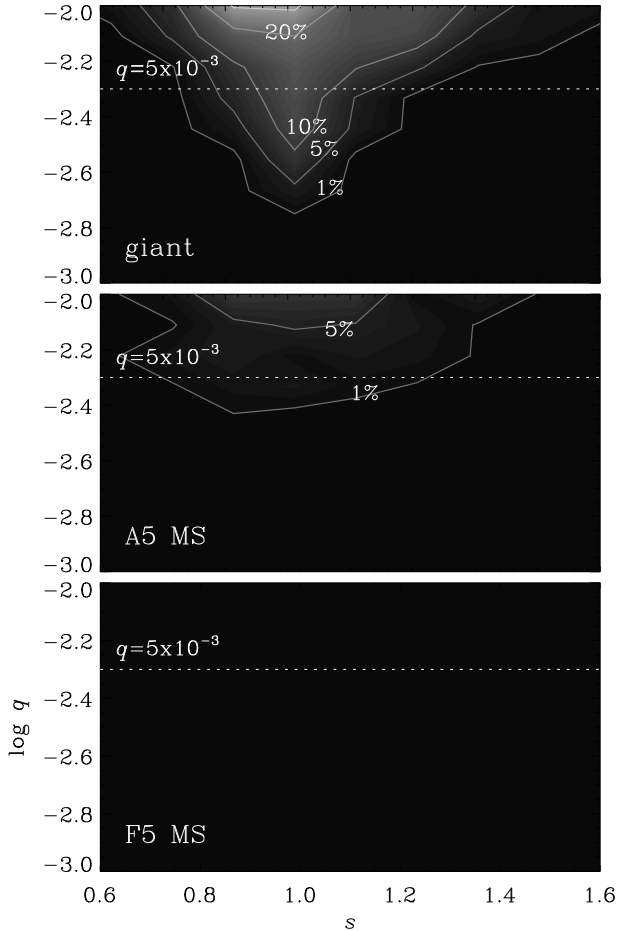}
\caption{\label{fig:six}
The planet detection efficiency of the Angstrom survey as a function of 
the star-planet separation $s$, and the planet/star mass ratio $q$, for
M31 lensing events associated with various source stars.  The type of the 
involved source star is marked at the bottom left corner of each panel.  
The field surface brightness is assumed to be $\mu=18.0\ {\rm mag}/
{\rm arcsec}^2$.  
}\end{figure}

% Figure 7 ------------------------------------------------------
\begin{figure}[t]
\epsscale{1.2}
\plotone{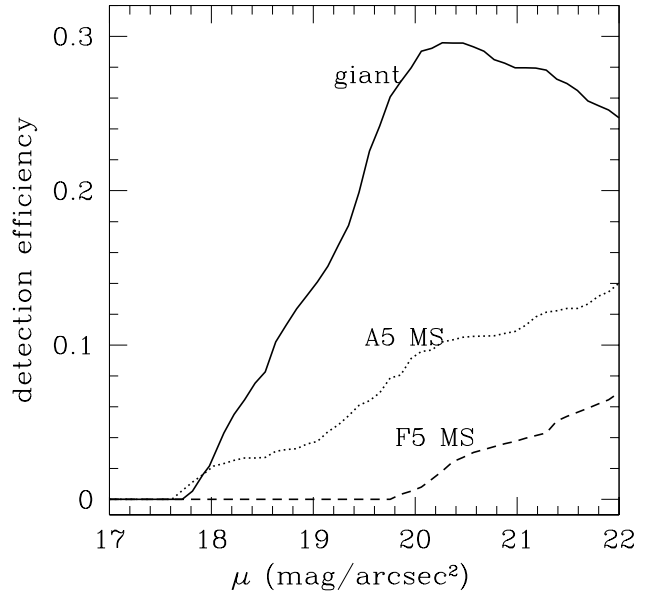}
\caption{\label{fig:seven}
Variation of the planet detection efficiency depending on the background
surface brightness $\mu$. The efficiency is estimated for the case of a
planetary system with the star-planet separation and the planet/star
mass ratio of $s=1.2$ and $q=5\times 10^{-3}$, respectively.
}\end{figure}

% Figure 8 ------------------------------------------------------
\begin{figure*}[t]
\epsscale{0.9}
\plotone{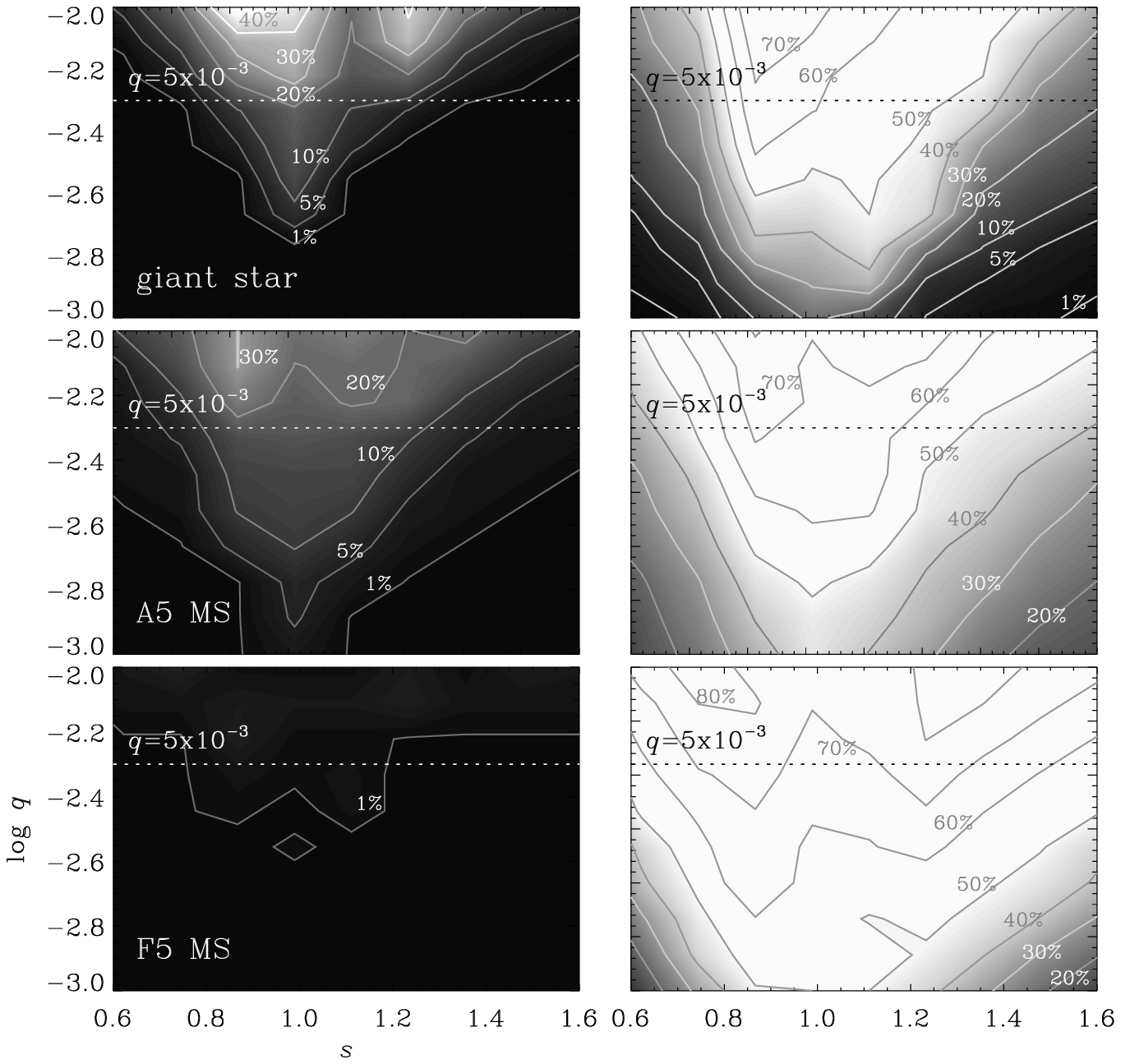}
\caption{\label{fig:eight}
Same as Fig.~\ref{fig:six}, but the efficiency is estimated under 
improved observational setups.  The left and right panels are the 
efficiencies estimated under the observational strategy II and III, 
respectively.  See the details of the observational strategies in \S\ 4.
}\end{figure*}

In Figure~\ref{fig:six}, we present the estimated detection efficiency
as a function of $s$ and $q$ for events involved with various source stars.
In Figure~\ref{fig:seven}, we also present the variation of the efficiency
depending on the background surface brightness, where the efficiency is
estimated by varying $\mu$ but fixing the lens parameters as $s=1.2$ and
$q=5\times 10^{-3}$.  From the figures, we find the following results.
\begin{enumerate}
\item
For events associated with giant source stars, it will be possible to 
detect planets with masses equivalent to or heavier than the Jupiter 
($q\sim 3\times 10^{-3}$).  Although the efficiency varies considerably 
depending on the star-planet separation, the average efficiency is 
$\sim 3\%$ for events caused by a lens system having a planet with a mass 
ratio $q=5\times 10^{-3}$ and located in the lensing zone.  However, it 
is expected that detecting planets with masses less than the Saturn 
($q\sim 10^{-3}$) would be difficult.

\item
The optimal events for the detections of the planetary signals are those
associated with giant stars.  It is expected that detecting planets for 
events associated with MS stars would be difficult because of the poor 
photometry and the resulting short durations of the planetary signals.  
For example, the duration of the event associated with an F-type MS star 
is $t_{\rm dur} \leq 2 u_{0,{\rm th}} t_{\rm E}\sim 1\ {\rm day}$.  
Considering that the planetary perturbation region with $D\geq 3$ occupies 
a fraction of the detection zone (as shown in Fig.~\ref{fig:three}), the 
planetary signal would be too short for detection.  Although planets can 
be detected with a non-negligible efficiency ($\sim 1\%$) for events 
associated with A-type MS stars, the number of planet detections from 
these events would be small because of the rarity of early-type MS stars 
projected on the M31 bulge region.  MS-associated events could be detected 
at low surface-brightness regions as shown in Figure~\ref{fig:seven}, but 
the event rate toward these fields would be low due to the low column 
density of lens matter along the line of sight.

\item
The efficiency peaks at a certain surface brightness.  In the region 
of very high surface brightness, planet detection is limited by the 
poor photometry.  In the very low surface-brightness regime, on the 
other hand, the event detection zone is significantly larger than the 
size of the planetary deviation region, which is confined around caustics.  
As a result, the efficiency, which is proportional to the one-dimensional 
size ratio of the planetary deviation region to the detection zone, is 
low in this region.  The peak efficiency occurs at 
$\mu\sim 20\ {\rm mag}/ {\rm arcsec}^2$ for events involved with giant 
source stars (see Figure~\ref{fig:seven}).

\item
In most cases, the planetary perturbations of M31 pixel-lensing events 
are induced by central caustics. Therefore, an observational strategy of 
focusing on these perturbations would maximize the detections of M31 
planets.  An alert system based on real-time survey observations combined 
with prompt follow-up observations would do this.

\end{enumerate}

\begin{deluxetable}{ccc}
\tablecaption{Detection Efficiencies Under Various Strategies \label{table1}}
\tablewidth{0pt}
\tablehead{
%\multicolumn{1}{c}{observational} &
%\multicolumn{3}{c}{detectability} \\
\colhead{strategy} &
\colhead{source star type} &
\colhead{detection efficiency} }
\startdata
I   & giant\ \ \ \     & 3\%   \\
    & A5 MS    & 1\%   \\
\smallskip
    & F5 MS    & 0\%   \\

II  & giant\ \ \ \     & 7\%   \\
    & A5 MS    & 9\%   \\
\smallskip
    & F5 MS    & 1\%   \\

III & giant\ \ \ \     & 41\%  \\
    & A5 MS    & 52\%  \\
    & F5 MS    & 66\%  \\
\enddata
\tablecomments{ 
Average detection efficiencies of detecting a planet with $q=5\times 10^{-3}$
located in the lensing zone of $0.6\leq s\leq 1.6$ under three different 
observational strategies.  The individual strategies imply a survey mode 
with a monitoring frequency $f=5\ {\rm times}/{\rm day}$ (strategy I), a 
survey mode with $f=10\ {\rm times}/{\rm day}$ (strategy II), and a survey 
mode with $f=5\ {\rm times}/{\rm day}$ plus follow-up observations by 
employing a single large telescope (strategy III), respectively. 
}
\end{deluxetable}

\section{Improving Planet Detection Efficiency}

Considering that planetary perturbations, in many cases, are missed 
from detection due to short durations, a significant improvement in 
the planet detection efficiency is expected with increased monitoring 
frequency.  One way to do this is using more telescope time or employing 
more telescopes (`strategy II'). The other way is conducting follow-up 
observations for events detected in the early phase by the survey 
experiment (strategy `III').  In this section, we estimate the efficiencies 
expected under these improved observational strategies.  We designate the 
observational condition of the current pixel-lensing survey (with 
$f=5\ {\rm times}/ {\rm day}$) as `strategy I'.

We simulate the observations under the strategy II by doubling the 
monitoring frequency of the current Angstrom experiment, i.e.\ 
$f=10\ {\rm times}/ {\rm day}$.  For strategy III, we assume survey-mode 
observations with $f=5\ {\rm times}/{\rm day}$ and follow-up observations 
with a single 8m-class telescope.  Follow-up observations are assumed to 
be carried out 4 hours after the first pixel-lensing event signal is 
detected by the survey observations and the assumed monitoring frequency 
is $f=20\ {\rm times}/ {\rm night}$.  Since a single telescope is employed, 
follow-up observations are assumed to be carried out only during the night 
(8 hrs per day), and thus 24 min per each combined image.  Assuming 20 min 
per exposure (to allow $\sim 4$ min for readout), the photometric uncertainty 
of the follow-up observation is $\sigma_{\rm 8m}/\sigma_{\rm 2m}\sim 
(30\ {\rm min}/20\ {\rm min})^{1/2} (2{\rm m}/8{\rm m})\sim 31\%$ of the 
survey observation.

In Figure~\ref{fig:eight}, we present the efficiency expected under the 
two improved observational strategies.  In Table~\ref{table1}, we also 
present the average detection efficiencies of detecting a planet with 
$q=5\times 10^{-3}$ located in the lensing zone under the three different 
observational strategies.  From the figure and table, we find that 
significant improvement in efficiency is expected, especially from the 
adoption of the follow-up observation strategy.  The improvement is more 
significant for events involved with MS stars because the short-duration 
perturbations associated with these events are readily detectable with 
the increased monitoring frequency.

\section{Conclusion}

We explored the feasibility of detecting planets in M31 from a
high-frequency pixel-lensing survey using a global network of 2m-class
telescopes.  For this evaluation, we estimated the efficiency of detecting
planetary signals for events induced by planetary systems with various
planet/star mass ratios and star-planet separations, associated with source
stars of various types, and occurring toward fields with various surface
brightness.  From the dependence of the detection efficiency on these
parameters, we found that 3$\sigma$ detection of the signals produced by
giant planets located in the lensing zone with masses equivalent to or
heavier than the Jupiter would be possible  with detection efficiencies
$\sim 3\%$ if the event is associated with giant stars.  A dramatic 
improvement of the efficiency is expected if follow-up observations based 
on real-time survey observations become possible.

\acknowledgments
Work by C.H. was supported by the Astrophysical Research Center for the
Structure and Evolution of the Cosmos (ARCSEC) of the Korea Science \&
Engineering Foundation (KOSEF) through the Science Research Program (SRC)
program.  B.-G.P. and Y.-B.J. acknowledge the support of the Korea Astronomy
and Space Science Institute (KASI).
E.J.K. was supported by an Advanced Fellowship from the UK Particle Physics 
and Astronomy Research Council (PPARC).  M.J.D. and J.P.D. were supported, 
respectively, by a PPARC post-doctoral research assistantship and PhD 
studentship.
A.G. was supported in part by grant AST 042758 from the NSF.  Any opinions, 
findings, and conclusions or recommendations expressed in this material 
are those of the authors and do not necessarily reflect the view of the NSF.

\end{document}